\documentstyle[aps,prl,multicol,epsf]{revtex}
\begin{document}
 \newcommand \be {\begin{equation}}
\newcommand \ee {\end{equation}}
 \newcommand \ba {\begin{eqnarray}}
\newcommand \ea {\end{eqnarray}}
\def\bphi{{\mbox{\boldmath{$\phi$}}}}
\newcommand{\subs}[1
]{{\mbox{\scriptsize #1}}}
\begin{multicols}{2}

\noindent{\large\bf Comment on ``Roughening Transition of Interfaces in
Disordered Media''}

\bigskip
Emig and Nattermann \cite{Emig} have recently investigated the competition between
lattice pinning and impurity pinning using the Functional Renormalisation
Group ({\sc frg}) approach. For elastic objects of internal dimensions $2 < D < 4$, they find, at
zero temperature, an interesting second order phase transition between a {\it flat} phase for
small disorder and a {\it rough} phase for large disorder. These results contrast with those obtained using the replica variational approach for the 
same problem, where a first order transition between flat and rough phases 
was predicted \cite{jpb}. In this comment, we show that these results can be reconciled
by analysing the {\sc frg} for an arbitrary dimension $N$ for the displacement
field ${\bphi}$. More specifically, we have considered the following Hamiltonian \cite{jpb}:
\ba \nonumber
{\cal H}[{\bphi}]&=&\int d^D\vec x \ \ \big[\frac{\gamma}{2}\sum_{i=1}^N \sum_{\alpha=1,D}
(\frac{\partial \phi_i}{\partial x_\alpha})^2\\
&+& V_\subs{imp}(\vec x,{\bphi})
+ v \ \cos\left(2\pi \sum_{i=1}^N \frac{\phi_i}{\sqrt{N}}\right)\big],
\ea
where $v$ measures the strengh of the periodic potential, acting along the
$(1,1,...1)$ direction and $V_\subs{imp}(\vec x,{\bphi})$ is a disordered Gaussian pinning 
field, such that:
\be
\overline{V_\subs{imp}(\vec x,{\bphi})V_\subs{imp}(\vec x',{\bphi'})}=
N W\delta^D(\vec x-\vec x') R({\bphi-\bphi'}).
\ee  
We have performed the {\sc frg} calculations using both the replica method and the dynamical method based on the Martin-Siggia-Rose action. Taking into account a direct contribution of $v$ in the renormalisation of $R({\bphi})$ (absent in \cite{Emig}), we find 
that {\it two} curvatures of $R$ must be introduced, according to:
$\partial_i \partial_j R(0) = -\Delta \delta_{ij} -\Delta'$. The coupled
flow equations read, in the large $\ell$ limit and for $\epsilon = 4-D$ small:
\ba\nonumber
\frac{d\Delta}{d\ell}&=&\epsilon \Delta - N \Gamma v^2 \Delta \\ \nonumber
\frac{d\Gamma}{d\ell}&=&\epsilon \Gamma -N(1-\frac{N}{4})v^2\Gamma^2 \\
\frac{dv}{d\ell}&=&(2-\frac{\Gamma}{2})v-N(\frac{1}{2}+\frac{3N}{8})\Gamma v^3
\ea
where $\Gamma=(\Delta+N\Delta')$, and some rescaling of $v$, $\Delta$ and $\Delta'$
has been performed (see \cite{Emig} for details). 
It is easy to show that for $N < 4$, a non trivial fixed point exists, given 
by: $\Delta^*=0$, $\Gamma^*=4-\epsilon(4+3N)/(4-N)$ and $v^*=\epsilon/N(4-N)$, with one unstable eigenvalue
$\lambda=2\sqrt{\epsilon}-\epsilon (4+N)/(4-N)$, which controls the divergence of the correlation length near the transition \cite{Emig}. The value for $\lambda$ agrees, to lowest order, with that obtained in \cite{Emig}. For $N > 4$, however, this fixed point disappears. We interpret this as a signal that 
the transition may indeed become first order for $N > 4$, 
as suggested by the variational approach of \cite{jpb} (which becomes exact for $N \to \infty$). The variational approach would therefore qualitatively 
correct for $N > 4$. 

The theory was extended in \cite{jpb,Emig} to treat the case of a non local 
elasticity, which corresponds to several situations of physical interest, such as
Bloch walls \cite{jpb,Emig} or contact lines \cite{Rollet,Anusha}. In this case, however,
the non local elastic term is {\it not} renormalized, and therefore {\it no 
fixed point} is found to order $\epsilon$ (contrarily to the claim of \cite{Emig}), suggesting again that the transition
is first order. The variational theory is thus also trustworthy in these cases. In this respect, it would be interesting to study contact lines 
on a disordered substrate, with a superimposed periodic modulation of wettability.
This case corresponds formally to a $D=2$ object with local elasticity, where the roughening transition becomes a crossover \cite{jpb}. The crossover length is however exponentially large in $1/W$ and might therefore not preclude the
observation of interesting effects.

\vskip 0.5cm
We thank P. Chauve, T. Emig, T. Nattermann and M. M\'ezard for useful discussions.

\bigskip
{\obeylines
\noindent Anusha Hazareesing\footnote{also at: Laboratoire de Physique Th\'eorique, E.N.S, 24 rue Lhomond, 75230 Paris {\sc cedex} 05.} and Jean-Philippe Bouchaud
CEA --- Service de Physique de l'Etat Condens\'e
Centre d'Etudes de Saclay
91191 Gif-sur-Yvette, France. 
}

\noindent
PACS numbers: 05.20.-y, 05.70.Jk

\end{multicols}

\end{document}